\documentclass[twoside,a4paper,12pt]{article}
\usepackage{graphicx}
\usepackage{dcolumn}
\usepackage{bm}
\usepackage{amsmath}
\usepackage{amsthm}
\usepackage{amsfonts}
\usepackage{times}
\usepackage{hyperref}
\usepackage{dsfont}
\usepackage{graphicx}
\usepackage{epsfig}
\usepackage[usenames, dvipsnames]{color}
\usepackage[noadjust]{cite}
\usepackage{authblk} 

\newtheorem*{theorem*}{Theorem}

\newtheorem*{remark*}{Remark}

\title{Cosmological  wave maps}
\author[1]{Spiros Cotsakis\footnote{On leave from the University of the Aegean}\footnote{\url{skot@aegean.gr}}}
\author[2]{John Miritzis\footnote{\url{imyr@aegean.gr}}}
\author[3]{Koralia Tzanni\footnote{\url{tzanni@aegean.gr}}}

\affil[1]{Department of Mathematics, American University of the Middle East, P.O. Box 220 Dasman, 15453, Kuwait}
\affil[2,3]{Department of Marine Sciences, University of the Aegean, University Hill, Mytilene 81100, Greece}

\begin{document}
\maketitle

\begin{abstract}
\noindent We consider theories of gravity that include many coupled scalar fields with
arbitrary couplings,  in the geometric  framework of wave maps. We examine the possibility
of obtaining acceptable cosmological solutions without the inclusion of a
potential term to the scalar fields. To illustrate the theory, we study two
simple models and compare their solutions to those in General Relativity. We also
address the issue of the conditions that must be satisfied by the wave maps for an accelerated phase of the Universe.

\end{abstract}

\section{Introduction}\label{intro}

The expansion history of the Universe is marked by two
characteristic phases, namely, inflation at the early stage and the present
longer period of acceleration. In both situations, scalar fields are essential
ingredients in the construction of cosmological scenarios aiming to describe
the evolution of the early and the present Universe. The accelerating phase of
the Universe requires scalar fields with non-negative potentials playing the
role of a cosmological term. Therefore, scalar fields have been a trustworthy
tool used by theorists for the explanation of the accelerating expansion of
the Universe \cite{wein2}.

Scalar fields arise in several conformally equivalent theories
of gravity, e.g., in higher order gravity theories \cite{baco88}, as well as in string
cosmology \cite{gasp}. The conformal potential which appears in the former class, \cite{baco88}, provides the best-fit, generic explanation to the 9-year WMAP and the Planck data  amongst all other options, when we set the degree of the polynomial lagrangian $n$ equal to half the spacetime dimensionality $D=2n$. An example of such a model belonging in the generic class $D=2n$ is the special case $D=4, n=2$ ($R^2$-inflation), cf. \cite{wmap}, Fig. 7 and Appendix, \cite{planck}, Fig. 8, where the latest Planck results \cite{planck} enhance their previous 2013 and 2015 results.

The simplest scalar-tensor theory developed by Brans and Dicke also 
involves a massless scalar field with constant coupling to matter, while
generalizations of the Brans-Dicke action led to scalar-tensor theories involving scalar field self-interactions and  dynamical couplings to matter \cite{maeda,fara}. For cosmological models based on scalar-tensor theories from a similar, dynamical systems perspective as the one studied in this work, see \cite{1,2,3,4,5}. Further
generalization is achieved by considering models with multiple scalar fields (see e.g.
\cite{daes}). For recent studies on multiple scalar field models see for
example \cite{cft,li}. Multifield coupled quintessence was investigated in
\cite{abn} in the framework of assisted dark energy or assisted quintessence.
For a detailed investigation on dark energy models using dynamical systems
methods see \cite{bbccft} and references therein.

A unified way to treat various aspects of scalar fields models interacting with each other is provided by the concept of a wave map. A wave map is a map from a spacetime to a Riemannian manifold. More
precisely, let $(\mathcal{M}^{m},g_{\mu\nu})$ be a spacetime, $(\mathcal{N}%
^{n},h_{ab})$ be a Riemannian (or semi-Riemannian) manifold and $\phi
:\mathcal{M}\rightarrow\mathcal{N}$ be a smooth map. $\mathcal{M}$ is called
the source manifold and $\mathcal{N}$ the target manifold.

Wave maps have been a focus of deep studies, and we know that they are regular in the spherically symmetric case under suitable conditions on the Riemannian target \cite{christ}, and also that they are stable for compactly supported initial data, remaining  close to spherical symmetry \cite{kr}.
Physical systems with multiple scalar fields can also be incorporated into the concept of suitable wave maps. The fundamental link of wave maps to multifield models is that  the
scalar fields $\phi^{a},a=1,\dots,n$ appear as the coordinates parameterizing the
Riemannian target. The metric $h_{ab}$ expresses the possible couplings between the
scalar fields. There are a number of basic advantages in treating interacting scalar fields as wave maps over the more widely used formulation:
\begin{itemize}
\item It provides an alternative, elegant,  invariant, coordinate-free treatment of any model that contains many scalar fields. This is clearer conceptually, and complementary to the usual, coordinate-based, intuitive  description.
\item It gives a more  economical way to describe the models, and so it unifies various seemingly unrelated properties of different scalar field models. If a property holds for the wave map, it is automatically valid for any multi-scalar field model obtained from it.
\item It offers the possibility to ask novel questions, especially global ones, very difficult to think in the standard approach. For example, how do the connections or the curvatures of the source or the target spaces of the wave map affect the global behaviour of the interacting scalar fields? More generally, what are the implications of various possible geometric or topological constructions related to a given wave map model for the resulting multiple, interacting scalar fields?
\item We may always revert to the coordinate description, if we wish to compare the invariant results obtained to those of the coordinate approach.
    \end{itemize}

In this work we examine the opposite case, that of theories without potentials for the scalar fields, where the scalar field interactions drive the evolution instead of their possible potentials. We imagine that throughout its history (or at least for the periods we focus in this paper), the Universe is driven by the dynamics of  a wave map with vanishing potential energy.  (For some results with a non-zero potential, see e.g., \cite{jarv1} using, however, a coordinate description). In the simplest case considered in the present work, we assume the target space to be one-dimensional (Section \ref{1D}), or two-dimensional with a diagonal metric (Section \ref{2D}). We believe that other choices of target, such as a hyperbolic space,  can lead to different results but this is in fact a far more complicated case.

The plan of the paper is as follows. In the next section we introduce the wave map model and prove  an equivalence of the resulting wave map-tensor theories to wave maps with
arbitrary couplings to the curvature. In Section \ref{appli}, we study the possibility of obtaining acceptable cosmological solutions
without imposing a potential term to the scalar fields. To this end, in Sections \ref{1D} and \ref{2D} we construct two specific models that incorporate wave maps, and compare the resulting wave map solutions to the corresponding
ones known in General Relativity. Finally, in Section \ref{disc} we further discuss the issue
under which conditions our model could predict accelerated expansion of the Universe.

\section{Einstein-wave map system}\label{wav}
We consider the action
\begin{equation}
S=\int_{\mathcal{M}}\sqrt{-g}dx\left(  R-g^{\mu\nu}h_{ab}\partial_{\mu}%
\phi^{a}\partial_{\nu}\phi^{b}\right)  , \label{eiwmaction}%
\end{equation}
where $R$ is the scalar curvature of the spacetime. We use Greek indices for
tensor fields on the source manifold $\mathcal{M},$ Latin indices for the
tensor fields on the target manifold $\mathcal{N}.$ Variation with respect to
the metric $g$ and the fields $\phi^{a},$ leads to the so called Einstein-wave
map system, namely
\begin{align}
G_{\mu\nu}  &  =h_{ab}\left(  \partial_{\mu}\phi^{a}\partial_{\nu}\phi
^{b}-\frac{1}{2}g_{\mu\nu}g^{\rho\sigma}\partial_{\rho}\phi^{a}\partial
_{\sigma}\phi^{b}\right)  ,\label{g}\\
\square\phi^{a}  &  +\Gamma_{bc}^{a}(h)g^{\mu\nu}\partial_{\mu}\phi
^{b}\partial_{\nu}\phi^{c}=0, \label{el}%
\end{align}
where $\square= g^{\mu\nu} \nabla_{\mu}\nabla_{\nu}$ and $\Gamma_{bc}^{a}(h)$
is the metric connection of the target manifold $\mathcal{N}.$ These equations
constitute a quasi-linear system of hyperbolic PDEs for $\phi^{a}$. The stress-energy tensor of the wave map, as defined in
the standard way through the basic wave map Lagrangian, is given by
\[
T_{\mu\nu}=h_{ab}\left(  \partial_{\mu}\phi^{a}\partial_{\nu}\phi^{b}-\frac
{1}{2}g_{\mu\nu}g^{\rho\sigma}\partial_{\rho}\phi^{a}\partial_{\sigma}\phi
^{b}\right)  .
\]
It has the properties of being symmetric, divergence-free and $T_{\mu\nu
}u^{\mu}u^{\nu}\geq0$ for all future-directed timelike vector fields $u$. In
the simple case $\mathcal{N}=\mathbb{R}$, we see that $T_{\mu\nu}$ is reduced
to the stress-energy tensor of a massless scalar field.


One may be tempted to generalize the theory (\ref{eiwmaction}) with arbitrary
couplings to the curvature and kinetic terms, i.e.,
\[
S=\int_{\mathcal{M}}\sqrt{-g}dx\left(  A(\phi)R-B(\phi)g^{\mu\nu}%
h_{ab}\partial_{\mu}\phi^{a}\partial_{\nu}\phi^{b}\right)  ,
\]
where $A,B$ are arbitrary smooth non-negative functions of $\phi$ (for further details see \cite{comi1}, \cite{comi2}). Such a
theory includes as special cases many of the scalar field models considered in
the literature (e.g., \cite{daes,barr,stac}.). If $B\left(  \phi\right)  $ is
a strictly positive function, it can be absorbed from the beginning into the
metric of the target manifold, i.e., without loss of generality we may set
$B=1$. Furthermore, defining a new metric by
\[
\tilde{g}_{\mu\nu}=A(\phi)g_{\mu\nu},
\]
the scalar curvature transforms as
\[
R=A\left(  \tilde{R}+3\square\ln A-\frac{3}{2}\tilde{g}^{\mu\nu}\frac
{\partial_{\mu}A\partial_{\nu}A}{A^{2}}\right)  .
\]
Dropping a total divergence and noting that $\partial_{\mu}A=\left(  \partial
A/\partial\phi^{a}\right)  \partial_{\mu}\phi^{a}$, the action becomes,
\[
\tilde{S}=\int_{\mathcal{M}}\sqrt{-\tilde{g}}dx\left(  \tilde{R}-\tilde
{g}^{\mu\nu}\left(  \frac{3}{2A^{2}}A_{a}A_{b}+\frac{B}{A}h_{ab}\right)
\partial_{\mu}\phi^{a}\partial_{\nu}\phi^{b}\right)  ,
\]
where $A_{a}=\partial A/\partial\phi^{a}$. In general, the quadratic form
\begin{equation}
\pi_{ab}:=\frac{3}{2A^{2}}A_{a}A_{b}+\frac{B}{A}h_{ab} \label{quad}%
\end{equation}
is not positive definite, unless one imposes further conditions on the
functions $A$ and $B.$ For example, in simple scalar-tensor theories, with
$A\left(  \phi\right)  =\phi$ and $B\left(  \phi\right)  =\omega\left(
\phi\right)  /\phi,$ one has to impose the condition that $\omega\left(
\phi\right)  \geq-3/2,$ in order that the energy density of the scalar field
be non-negative. Assuming that $\mathrm{rank\,}\pi_{ab}=\dim\mathcal{N},$ we
may define the reciprocal tensor $\pi^{ab},$ i.e., $\pi^{ab}\pi_{bc}%
=\delta_{c}^{a}.$ Therefore the target manifold is endowed with a new metric,
and we use $\pi_{ab}$ to raise and lower indices in $\mathcal{N}.$ Using eq.
(\ref{quad}), the original action becomes that of a wave map minimally coupled
to the Einstein term
\[
\tilde{S}=\int_{\mathcal{M}}\sqrt{-\tilde{g}}dx\left(  \tilde{R}-\tilde
{g}^{\mu\nu}\pi_{ab}\partial_{\mu}\phi^{a}\partial_{\nu}\phi^{b}\right)  .
\]
This result shows that under certain conditions, couplings of the wave map to
the curvature are conformally equivalent. If ordinary matter is present, then an extra term $S_{\rm{matter}}$ should be added to the action \eqref{eiwmaction}. For the possible invariant quantities associated with conformal transformations of multifield models (independently of the wave map perspective studied here), see \cite{jarv2}.

\section{Cosmological wave maps}\label{appli}

We consider flat, homogeneous and isotropic spacetimes and adopt the metric
and curvature conventions of \cite{wael}. Units have been chosen so that
$c=1=8\pi G$. The Hubble function is defined by $H\equiv\dot{a}/a$, where
$a\left(  t\right)  $ is the scale factor and an overdot denotes
differentiation with respect to time $t$. We assume that ordinary matter is
described by a perfect fluid with equation of state $p=(\gamma-1)\rho$, where
$0<\gamma<2$. The Bianchi identities imply that the total energy-momentum
tensor is conserved, and assuming that there is no energy exchange between the
perfect fluid and the scalar fields, each component is separately conserved.
The field equations reduce to the Friedmann equation,
\begin{equation}
3H^{2}=\rho+\frac{1}{2}h_{ab}\dot{\phi}^{a}\dot{\phi}^{b};\label{frie}%
\end{equation}
the Raychaudhuri equation,
\begin{equation}
\dot{H}=-\frac{1}{2}h_{ab}\dot{\phi}^{a}\dot{\phi}^{b}-\frac{\gamma}{2}\rho;\label{ray}
\end{equation}
the wave map equations \eqref{el},
\begin{equation}
\ddot{\phi}^{a}+3H\dot{\phi}^{a}+\Gamma_{bc}^{a}\dot{\phi}^{b}\dot{\phi}^{c}=0;\label{ems}
\end{equation}
and the conservation equation of the perfect fluid,
\begin{equation}
\dot{\rho}=-3\gamma\rho H.\label{conss}%
\end{equation}

A significant part of models studied so far in the literature, especially in string cosmology \cite{gasp}, depend on the choice of the target metric $h_{ab}$. For instance, if $\mathcal{N}=\mathbb{R}$ then $\phi$ is simply a real scalar field on the source $\mathcal{M}.$ If, on the other hand, we choose $\mathcal{N}=\mathbb{R}^{n}$ then the wave map $\phi=\left(  \phi_{1},...,\phi_{n}\right)  $ may be thought of as $n$ \textit{uncoupled} scalar fields on $\mathcal{M}$.

We begin with some general properties of the system \eqref{ray}--\eqref{conss} with the constraint \eqref{frie}. Firstly, the system shares the remarkable property of the Einstein equations that, if equation (\ref{frie}) is satisfied at some initial time, then it is satisfied throughout the evolution. We recall that $h_{ab}$ is positive definite, i.e., $h_{ab}\dot{\phi}^{a}\dot{\phi}^{b}>0$, for all nonzero scalar field velocities $\dot{\phi}^{a}$. Secondly, following the arguments in \cite{fost, miri03}, one can show that an initially expanding flat universe remains ever-expanding. In fact, the set $\{ (\phi^{i}, \dot{\phi}^{i},\rho,H),~H=0 \}$ is invariant under the flow of the system \eqref{ray}--\eqref{conss} with the constraint \eqref{frie}. Therefore, the sign of $H$ is
invariant. If the sign of $H$ could change, a solution curve starting with say, a positive $H$, would pass through the origin which is an equilibrium solution of \eqref{ray}--\eqref{conss}, thus violating the fundamental existence and uniqueness theorem of differential equations. We conclude that initially expanding universes remain ever expanding. Thirdly, equation \eqref{conss} of the above system implies that also the set $\rho=0$ is invariant. Therefore, if initially $\rho$ is positive, it remains positive for ever. Furthermore, for $\rho>0$ the second equation \eqref{ems} of the system implies $\dot{H}<0$, thus $H$ is a decreasing function of time $t$.

In the following Sections, we construct two examples to illustrate the theory.

\section{One-dimensional target manifold}\label{1D}

The simplest example one can construct is by considering a one-dimensional
target manifold. We assume that the metric has the form
\begin{equation}
h(\phi)=h_{0}e^{-2\lambda\phi}, \label{metric1D}
\end{equation}
where $h_{0}$ and $\lambda$ are positive, otherwise arbitrary constants.
The assumption that $\lambda$ is positive is not restrictive. For $\lambda=0,$ the target manifold is simply $\mathbb{R}$ and the system describes models with a massless scalar field. For negative $\lambda$, it will be clear at the end of this section that the nature of the equilibrium solutions remains the same and the corresponding phase portrait is symmetric with respect to the $y$ axis.

The
field equations \eqref{ray}--\eqref{conss} take the form
\begin{align}
3H^{2}=  &  \rho+\frac{h_{0}}{2}\dot{\phi}^{2}e^{-2\lambda\phi},\label{sys:1}%
\\
\dot{H}=  &  -\frac{h_{0}}{2}\dot{\phi}^{2}e^{-2\lambda\phi}-\frac{\gamma}%
{2}\rho,\label{sys:2}\\
\ddot{\phi}=  &  -3H\dot{\phi}+\lambda\dot{\phi}^{2},\label{sys:3}\\
\dot{\rho}=  &  -3\gamma\rho H. \label{sys:4}%
\end{align}

Equations \eqref{sys:2}--\eqref{sys:4} constitute a four-dimensional dynamical system
subject to the constraint \eqref{sys:1}.
We introduce expansion-normalized variables \cite{wael,clw}, defined as
\[
x=\frac{\dot{\phi}}{H}, \qquad y=\sqrt{\frac{h_{0}}{6}}e^{-\lambda \phi}x, \qquad \Omega=\frac{\rho}{3H^{2}}.
\]
We assume that initially the Universe is expanding. As discussed in the
previous section, it follows that $H>0$ for all $t$, thus the above new
variables are well defined.
Note that for $H>0$, the scale factor $a(t)$ is a monotone increasing function irrespective of the particular form of the metric $h_{ab}$. Therefore, we may define a new time variable as $\tau=\ln a$. Since $d\tau/dt>0$, the new time is also well defined.  In the following, a prime denotes differentiation with respect to the new time $\tau$.

An advantage of the expansion-normalized variables' formalism is that the evolution equation for $H$,
\[
H^{\prime}=H\left(  -3+\alpha  \Omega\right),  \qquad\alpha:=\tfrac{3}%
{2}\left(  2-\gamma\right)  >0,
\]
decouples from the rest of the evolution equations. In fact, the system becomes three-dimensional
\begin{align}
x^{\prime}=  &  \lambda x^{2}-\alpha x\Omega, \label{x:1}\\
y^{\prime}=  &  -\alpha y\Omega,\label{y:1}\\
\Omega^{\prime}=  &  2\alpha\Omega\left(  1-\Omega\right)  , \label{W:1}%
\end{align}
subject to the constraint
\begin{equation}
\Omega+y^{2}=1. \label{constr:1}%
\end{equation}
The physical values for $\Omega$ are, $\Omega\geq0$ and from the constraint we
get $\Omega\leq1$.

We observe that Eqs \eqref{y:1} and \eqref{W:1} can be written as a single
differential equation
\begin{equation}
\frac{dy}{d\Omega}=\frac{1}{2}\frac{y}{\Omega-1}, \label{ydif}
\end{equation}
with solution given by,
\begin{equation}
y=c\sqrt{1-\Omega}, \text{ for all } ~\Omega \in [0,1],~c\in \mathbb{R}.\label{y:branch}
\end{equation}
Note that the differential equation (\ref{ydif}) is defined for all $\Omega\neq1$, but
the solution (\ref{y:branch}) can be extended to the whole interval $\left[
0,1\right]  $.
According to the
sign of $c$, the graph of the function $y(\Omega)$, given in \eqref{y:branch},
has two branches. From Eq. \eqref{y:1} we conclude that the plane $y=0$ is an
invariant set, therefore the sign of $y$ is invariant. Hence $y$ stays on only
one branch of \eqref{y:branch} and thus takes either only positive or only
negative values. We arrive at the same conclusion by the continuity properties
of $y$.
From the definition of $y$, for expanding universes, $H>0$, a positive sign
of $y$ corresponds to $\dot{\phi}>0$, hence to an increasing function of $\phi $. Since the sets $y>0$ and $y<0$ are invariant we conclude that the sign
of $\dot{\phi}$ remains invariant and therefore the scalar field does not
oscillate.

A further reduction of the dimension of our dynamical system can be obtained
by eliminating $\Omega$ from equations \eqref{x:1}--\eqref{W:1}, using the
constraint \eqref{constr:1}. The system becomes
\begin{align}
x^{\prime}=  &  \lambda x^{2}+\alpha x(y^{2}-1),\label{x:2}\\
y^{\prime}=  &  \alpha y(y^{2}-1). \label{y:2}%
\end{align}
From \eqref{constr:1} the phase space of the system is given by the strip,
\begin{equation}
y^{2}-1\leq0, \qquad  x \in\mathbb{R}.
\end{equation}

\begin{figure}[th]
\begin{center}
\includegraphics[scale=0.3]{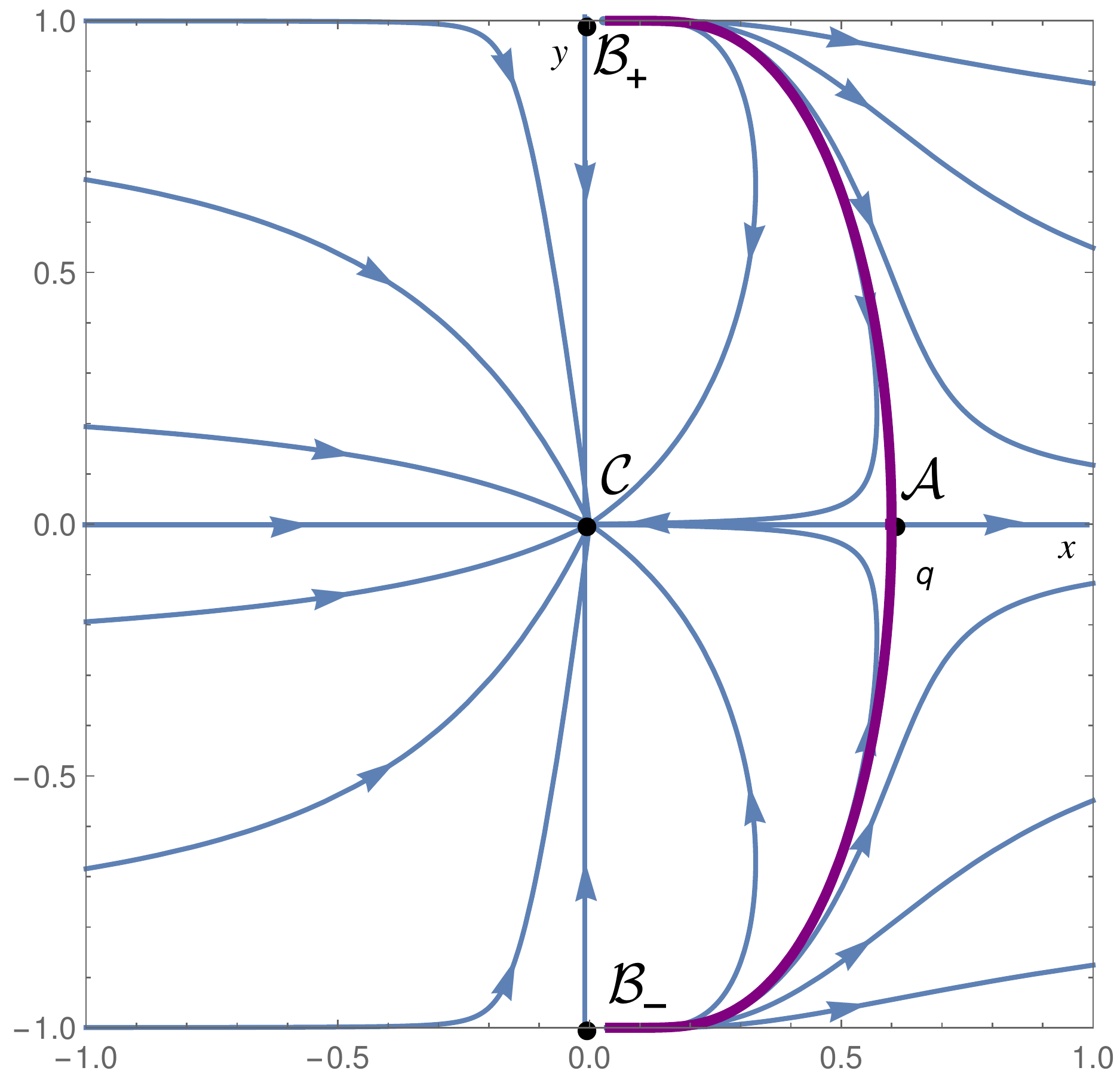}
\end{center}
\caption{Phase portrait of the system \eqref{x:2}--\eqref{y:2} for $\gamma=1$ and $\lambda=2.5$.}%
\label{fig1d}%
\end{figure}

The equilibrium points of the system \eqref{x:2}--\eqref{y:2} are shown in
Table \ref{table:1}. For the construction of the phase portrait we note that
the system \eqref{x:2}--\eqref{constr:2} is symmetric with respect to the $x$
axis, i.e. it is invariant under the transformation $\left(  t,y\right)
\rightarrow\left(  -t,-y\right)  $. That means that if $\left(  x\left(
t\right)  ,y\left(  t\right)  \right)  $ is a solution, then also $\left(
x\left(  -t\right)  ,-y\left(  -t\right)  \right)  $ is a solution and
therefore, every trajectory has a twin, symmetric with respect to the $x$ axis
and with opposite orientation. Eq. \eqref{x:2} implies that the solution
$\mathcal{C}$ attracts almost all trajectories starting to the left of the
separatrix $q$. Trajectories starting to the right of the separatrix $q$
diverge, approaching the limiting state of an ever-expanding Universe with
$H\rightarrow0$ and $\Omega\rightarrow1$. This is due to the fact that the
scalar field increases without bound and its contribution through
(\ref{metric1D}) to the evolution of the system fades. The phase portrait
of the system is shown in Figure \eqref{fig1d}.
\begin{table}[ptb]
\caption{Equilibrium Points of the system \eqref{x:2}--\eqref{y:2}}%
\label{table:1}
\begin{center}%
\begin{tabular}
[c]{llllll}\hline\hline
Label & $(x,y)$ & $\Omega$ & $a(t)$ & Stability & Acceleration\\\hline
$\mathcal{A}$ & $\left(  \alpha/ \lambda,0\right)  $ & $1$ & $t^{2/3\gamma}$ &
Saddle & $\gamma<2/3$\\
$\mathcal{B_{\pm}}$ & $\left(  0,\pm1\right)  $ & 0 & $t^{1/3}$ & Unstable &
Never\\
$\mathcal{C}$ & $\left(  0,0\right)  $ & 1 & $t^{2/3\gamma}$ & Stable &
$\gamma<2/3$\\\hline\hline
\end{tabular}
\end{center}
\end{table}

For the interpretation of the solutions let us stay at the upper half plane
$0\leq y\leq1$. Point $\mathcal{B}_{+}$ is an unstable past attractor and
represents a field dominated state. This explains the fact that near big bang
the scale factor evolves as $t^{1/3}$ (Dirac's cosmology \cite{wein}). On the contrary, points $\mathcal{A}$
and $\mathcal{C}$ are matter dominated and the scale factor evolves as in GR.
In fact, $y\geq0$ implies that $\phi\left(  t\right)  $ is an increasing
function of time and therefore the significance of the scalar field at late
times is negligible. Point $\mathcal{A}$ is a saddle and represents a matter
era with kinetic terms of the scalar field which do not contribute to the
total energy density. Point $\mathcal{C}$ is a future attractor and matter
dominated. Two conclusions can be drawn from the above analysis: First, there
exist no scaling equilibrium solutions and second, no equilibrium may
represent an accelerating phase of the Universe, except in the unrealistic
case $\gamma<2/3$.

\section{Two-dimensional target manifold}\label{2D}

\label{appli2}

A more complicated model can be constructed by a two-dimensional target
manifold. We assume that the metric is diagonal with
\begin{equation}
h_{11}\left(  \phi_{1},\phi_{2}\right)  =h_{0}e^{-2\lambda\phi_{1}-2\mu
\phi_{2}},\ \ h_{22}\left(  \phi_{1},\phi_{2}\right)  =\delta h_{0}%
e^{-2\lambda\phi_{1}-2\mu\phi_{2}},\label{hmet}%
\end{equation}
$\lambda,~\mu>0$. The constants $h_{0}$ and $\delta$ are assumed to be
positive which guarantees that the metric is positive definite. $\delta$ is a
measure of the difference of the ratio $h_{22}/h_{11}$ from unity. With this
choice all the connection coefficients $\Gamma$ of the target manifold are
constants, depending only on the parameters $\lambda,\mu$ and $\delta$.
Equations (\ref{ray})--(\ref{conss}) reduce to a six-dimensional dynamical
system
\begin{align}
\dot{H} &  =-\frac{1}{2}\left(  h_{11}\dot{\phi}_{1}^{2}+h_{22}\dot{\phi}%
_{2}^{2}+\gamma\rho\right)  ,\label{H:2}\\
\ddot{\phi}_{1} &  =-3H\dot{\phi}_{1}+\lambda\dot{\phi}_{1}^{2}+\mu\dot{\phi
}_{1}\dot{\phi}_{2}-\delta\lambda\dot{\phi}_{2}^{2},\label{f1:2}\\
\ddot{\phi}_{2} &  =-3H\dot{\phi}_{2}+\mu\dot{\phi}_{2}^{2}+\lambda\dot{\phi
}_{1}\dot{\phi}_{2}-\frac{1}{\delta}\mu\dot{\phi}_{1}^{2},\label{f2:2}\\
\dot{\rho} &  =-3\gamma\rho H,\label{rho:2}%
\end{align}
subject to the constraint
\begin{equation}
3H^{2}=\rho+\frac{1}{2}\left(  h_{11}\dot{\phi}_{1}^{2}+h_{22}\dot{\phi}%
_{2}^{2}\right)  .\label{constr:2}%
\end{equation}
Only Eq.\eqref{H:2} contains the fields $\phi_{1}$ and $\phi_{2}$ explicitly.
Therefore, using the constraint \eqref{constr:2} to eliminate the source term
$h_{ab}\dot{\phi}^{a}\dot{\phi}^{b}$ and noting that equations \eqref{f1:2}
and \eqref{f2:2} do not contain the fields $\phi_{1}$ and $\phi_{2}$
explicitly, we conclude that the system \eqref{H:2}--\eqref{rho:2} is actually
4-dimensional. A further reduction of the dimension is achieved by using new
variables defined as
\[
x=\frac{\dot{\phi}_{1}}{H},~~y=\frac{\dot{\phi}_{2}}{H},~~\Omega=\frac{\rho
}{3H^{2}},
\]
and a new time variable $\tau=\ln a$. Our system becomes
\begin{align}
x^{\prime} &  =\lambda x^{2}+2\mu xy-\delta\lambda y^{2}-\alpha x\Omega
,\qquad\alpha:=\tfrac{3}{2}\left(  2-\gamma\right)  >0\nonumber\\
y^{\prime} &  =\mu y^{2}+2\lambda xy-\frac{\mu}{\delta}x^{2}-\alpha
y\Omega,\label{sys3d}\\
\Omega^{\prime} &  =2\alpha\Omega\left(  1-\Omega\right)  ,\nonumber
\end{align}
where a prime denotes differentiation with respect to $\tau$. The evolution of
the Hubble function is described by the equation
\begin{equation}
H^{\prime}=H\left(  -3+\alpha\Omega\right)  ,\label{Hprime2}
\end{equation}
which decouples again from the rest of the evolution equations.

The equilibrium points are shown in Table \ref{tbl:2d}. Point $\mathcal{O}$
represents an ever-expanding Universe and is unstable. In fact, since
$\alpha>0 $ the energy density $\Omega$ in (\ref{sys3d}) satisfies the
logistic equation, therefore the plane $\Omega=0$ is unstable. We conclude that all solutions starting at any plane $0<\Omega<1$ eventually approach the plane $\Omega=1$.

\begin{figure}[th]
\begin{center}
\includegraphics[scale=0.3]{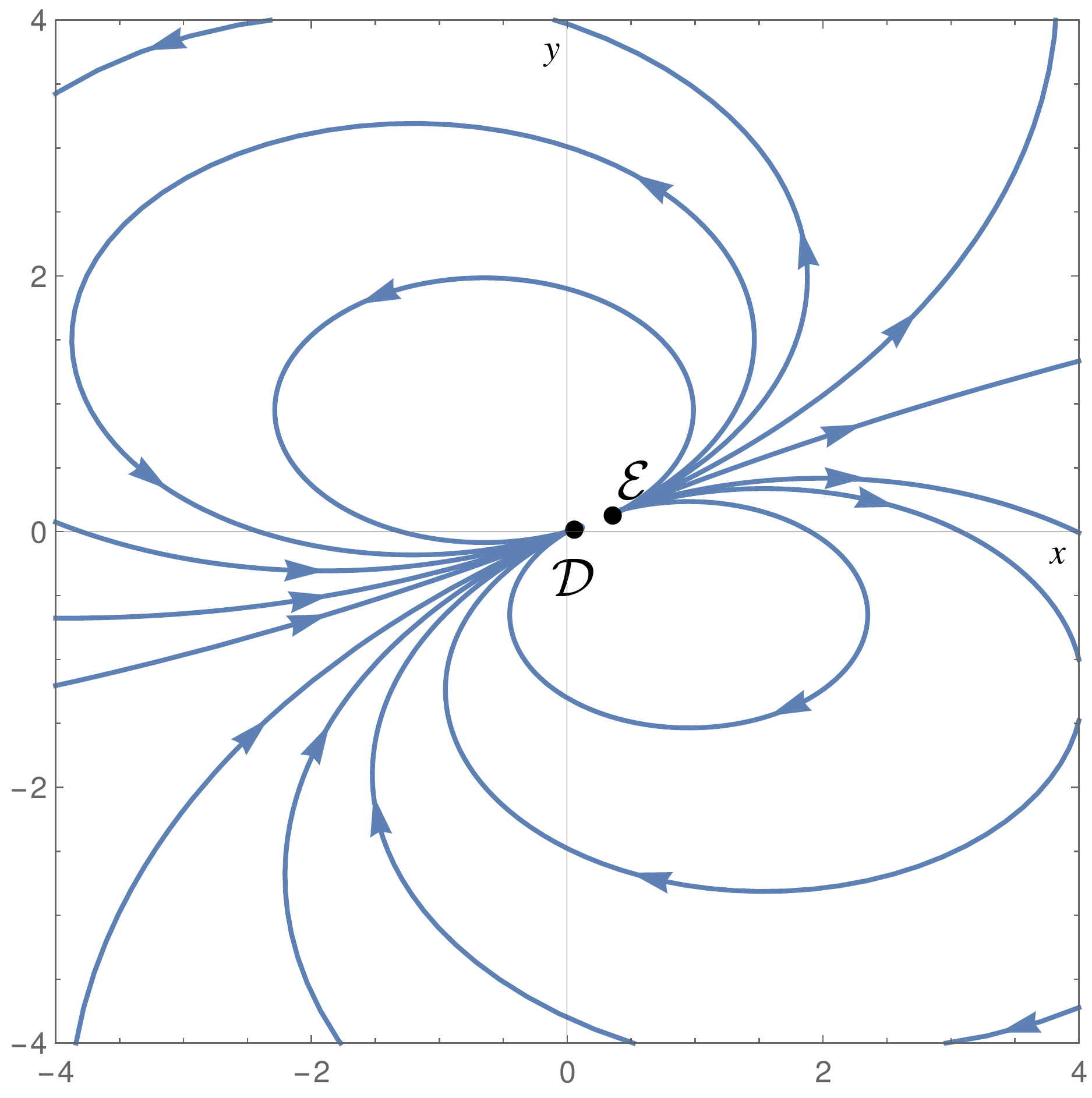}
\end{center}
\caption{Phase portrait of the projected three-dimensional system
\eqref{sys3d} on the invariant set $\Omega=1$, for the case where $\lambda
=\mu=\delta=2.5$ and $\gamma=1$.}%
\label{fig2d}%
\end{figure}

It is legitimate therefore to consider the projection of the system \eqref{sys3d} on that plane. The phase
portrait of this system is shown in Figure \ref{fig2d}.  The matter point $\mathcal{E}$ and late
attractor $\mathcal{D}$ both lie on the invariant plane $\Omega=1$. Point
$\mathcal{E}$ is of saddle type and trajectories passing nearby eventually
diverge. Note that Figure \ref{fig2d} does not exhibit the true character of solutions passing near $\mathcal{E}$, because only trajectories lying on the plane $\Omega=1$ are shown. Point $\mathcal{E}$ is a transient solution and although the kinetic terms are non zero the scalar fields themselves diverge to infinity in such a way that ordinary matter dominates. Point $\mathcal{D}$ is a stable node and represents a matter dominated phase. In all equilibrium points the scale factor does not depend on the explicit form of $h_{11},h_{22}$, because the effecive equation of state, $w_{\mathrm{eff}}:=-1-2\dot{H}/\left(3H^{2}\right)$, which determines the time dependence of the scale factor, (see Eq. \eqref{Hprime2}), is a function only of $\Omega$. Note that the equilibrium solutions
are the same as those obtained in GR. As in the one dimensional target, we observe that there
exist no scaling equilibrium solutions and also, except in the unrealistic
case $\gamma<2/3$,  no equilibrium may
represent an accelerating phase of the Universe.

\begin{table}[ptb]
\caption{Equilibrium Points in 2D}%
\label{tbl:2d}
\begin{tabular}
[c]{llllll}\hline\hline
Label & $(x,y,\Omega)$ & $a\left(  t\right)  $ & Eigenvalues & Stability &
Acceleration\\\hline
$\mathcal{O}$ & $\left(  0,0,0\right)  $ & $t^{1/3}$ & $2\alpha,0,0$ &
Unstable for all $\gamma$ & Never\\
$\mathcal{D}$ & $\left(  0,0,1\right)  $ & $t^{2/3\gamma}$ & $-\alpha
,-\alpha,-2\alpha$ & Stable for all $\gamma$ & $\gamma<2/3$\\
$\mathcal{E}$ & $\left(  \frac{\alpha\delta\lambda}{\delta\lambda^{2}+\mu^{2}%
}, \frac{\alpha\mu}{\delta\lambda^{2}+\mu^{2}},1\right)  $ & $t^{2/3\gamma}$ &
$\alpha,\alpha,-2\alpha$ & Saddle for all $\gamma$ & $\gamma<2/3$%
\\\hline\hline
\end{tabular}
\end{table}

\section{Discussion}\label{disc}

Wave maps provide an elegant formalism to describe many interacting scalar
fields. Our analysis was focused on the cosmological implications of theories
including many coupled scalar fields and a perfect fluid. The models developed in the previous sections share the same qualitative properties, namely
their equilibrium solutions correspond to those obtained in General Relativity. In particular,
the late attractors of both models are matter dominated, ($\Omega=1$), and
the scale factor evolves as in the known solutions in General Relativity. Thus, the Universe at late times evolves
ignoring any effect of the scalar fields. Therefore such systems cannot
predict acceleration. This is due to the absence from our model of a
cosmological term, such as a potential with nonnegative minimum. Moreover, the
positive values of $h_{ab}$ leave no room for a phantom regime. We emphasize
that our results in Section \ref{2D} do not depend on the functional form of
the metric coefficients of the target two-dimensional manifold. This is
reflected to the fact that the scale factor at the equilibrium points is
independent of $h_{11}, h_{22}$.

The above results are typical for the wave map system \eqref{g}--\eqref{el}
and do not depend either on the particular form of the metric $h_{ab}$ or on
the dimension of the target manifold. In fact, for the system \eqref{g}--\eqref{el} the Friedmann equation reads,
\begin{equation}
3H^{2}=T_{00}, \label{fried}%
\end{equation}
where the 00-component of the wave map energy-momentum tensor is given by,
\begin{equation}
T_{00}=h_{ab}\dot{\phi}^{a}\dot{\phi}^{b}. \label{ooen}%
\end{equation}
One expects that at critical points of $T_{00}$ the Universe inflates
exponentially. However, for semi-positive definite metric $h_{ab},$ it can be
shown that the only minima of $h_{ab}\dot{\phi}^{a}\dot{\phi}^{b}$ are zero.
Furthermore, for a positive definite metric $h_{ab},$ the minima of
$h_{ab}\dot{\phi}^{a}\dot{\phi}^{b}$ occur necessarily at $\dot{\phi}^{a}=0,$
$a=1,2,...,n$, i.e. equilibrium points of the system (\ref{g}) and (\ref{el}).
Therefore, equation (\ref{fried}) has no solutions with a constant nonnegative
$H$.

Of course, we could relax our assumption of positive definite metrics $h_{ab}
$, and consider the case of a semi-Riemannian target manifold. For example,
one may examine the two-dimensional metric (\ref{hmet}) with negative $\delta
$. In that case it is shown that the late attractor and all the transient
solutions, i.e. saddles, of the corresponding system (\ref{sys3d}) are again
matter dominated. Thus the simple change of the sign of $h_{22}$ does not produce acceleration. The construction of particular forms of non positive
definite metrics of the target manifold which allow for solutions with a
constant positive $H$, is the subject of future research.


\begin{thebibliography}{99}

\bibitem{wein2} Weinberg, S.:  Cosmology, Oxford University Press (2008)

\bibitem{baco88} Barrow, J. D., Cotsakis, S.: Phys. Lett B \textbf{214} 515 (1988)
\bibitem{gasp} Gasperini, M.: String Cosmology, Cambridge University Press (2007)
\bibitem{wmap}Hinshaw, G.F., et.al., 2013, ApJS., 208, 19H; arXiv: 1212.5226.

\bibitem{planck}The Planck Collaboration, \emph{Planck 2018 results. X. Constraints on inflation},   arXiv:1807.06211


\bibitem{maeda} Fujii, Y., Maeda, K. I.:  The scalar-tensor theory of gravitation, Cambridge University Press (2003)

\bibitem{fara} Faraoni, V.: Cosmology in scalar-tensor gravity, Springer Science \& Business Media (2004)

\bibitem{1} Romero, C., Oliveira, H. P., Mello Neto, J. R. T.: Astrophysics and space science \textbf{158} 229 (1989)

\bibitem{2} Kolitch, S.J., Eardley, D. M.: Ann. Phys. \textbf{241}  128 (1995)

\bibitem{3} Holden, 	D.J., Wands, D.: Class. Quant. Grav. \textbf{15}  3271 (1998)

\bibitem{4}	J\"{a}rv, L., Kuusk,  P., Saal, M.: Phys. Rev. D \textbf{78}  083530 (2008)

\bibitem{5} Coley,	A.A.: Gen. Rel. Grav. \textbf{31}  1295 (1999)

\bibitem {daes} Damour, T., Esposito-Far\`{e}se, G.: Class. Quantum Grav. \textbf{9}, 2093 (1992)

\bibitem {cft} Chiba, T., De Felice, A., Tsujikawa, S.: Phys. Rev. D \textbf{90}, 023516 (2014)

\bibitem {li} Li, Y.: Int. J. Mod. Phys. D \textbf{26}, 1750164 (2017)

\bibitem {abn} Amendola, L., Barreiro, T., Nunes, N.J.: Phys. Rev. D. \textbf{90} 083508 (2014)

\bibitem {bbccft} Bahamonde, S., Boehmer, C. G., Carloni, S., Copeland, E. J., Fang, W., Tamanini, N.: Physics Reports \textbf{775}  1 (2018)

\bibitem{christ} Christodoulou, D., Tahvildar-Zadeh, A. S.:  Communications on Pure and Applied Mathematics, \textbf{46}, 1041 (1993)

\bibitem{kr} Krieger, J.: Stability of spherically symmetric wave maps (No. 853), American Mathematical Soc. (2006)

\bibitem{jarv1} J\"{a}rv, L., Saal, M.,  Kuusk, P.:  Phys. Rev. D \textbf{81} 104007 (2010)


\bibitem{comi1} Cotsakis, S., Miritzis, J.: Proceedings of the IX Marcel Grossmann Meeting, On Recent Developments in Theoretical and Experimental General Relativity, Gravitation and Relativistic Field Theories, Rome 2000, V. Gurzadyan, R. T. Jantzen, R. Ruffini (eds), World Scientific pp 1945-1946, (2002), arXiv:gr-qc/0011086

\bibitem{comi2} Cotsakis, S., Miritzis, J.: Modern Theoretical and Observational Cosmology, pp 67-73, S. Cotsakis and M. Plionis (eds), Kluwer (2002), arXiv:gr-qc/0107100

\bibitem {barr}Barrow, J.D.: Phys. Rev. D \textbf{47}, 5329 (1993)

\bibitem {stac} Steinhardt, P.J., Accetta, F.S.: Phys. Rev. Lett. \textbf{64}, 2740 (1990)

\bibitem{jarv2}  J\"{a}rv, L., Kuusk, P., Saal, M., Vilson, O.: Phys. Rev. D \textbf{91} 024041 (2015)

\bibitem {wael} Wainwright, J., Ellis, G.F.R.: Dynamical Systems in Cosmology, Cambridge University Press (1997)



\bibitem{fost} Foster, S.: Class. Quant. Grav. \textbf{15}, 3485 (1998)

\bibitem {miri03} Miritzis, J.: Class. Quantum Grav. \textbf{20} 2981 (2003)

\bibitem {clw} Copeland, E. J., Liddle,  A. R., Wands, D.: Phys. Rev. D \textbf{57}, 4686 (1998)

\bibitem{wein} Weinberg, S.: Gravitation and Cosmology, Wiley (1972)


\end{thebibliography}
\end{document}